\documentclass[aps,prb,twocolumn,showpacs,preprintnumbers,superscriptaddress,floatfix]{revtex4-1}
\usepackage{tikz}
\usepackage{graphicx}
\usepackage{dcolumn}
\usepackage{float}
\usepackage{bm}
\usepackage{pifont}
\usepackage{cancel}
\usepackage{latexsym}
\usepackage{textcomp}
\usepackage{array}
\usepackage{amssymb}
\usepackage{amsfonts}
\usepackage{colortbl}
 \usepackage{array,multirow,graphicx}
\usepackage{afterpage}
\usepackage{amsmath}
\usepackage{fancybox}
\usepackage{color}
\usepackage{afterpage}
\usepackage{ulem}
\usepackage{textcase}
\usepackage[printonlyused]{acronym}
\definecolor{MyDarkBlue}{rgb}{0,0.08,0.45}
\newcolumntype{X}{>{\setbox0=\hbox\bgroup}c<{\egroup}@{}}

\newcommand{\cubra}[1]{\left\{#1\right\}}
\newcommand{\robra}[1]{\left(#1\right)}

\newcommand{\be}{\begin{equation}}
\newcommand{\ee}{\end{equation}}
\newcommand{\bea}{\begin{eqnarray}}
\newcommand{\eea}{\end{eqnarray}}
\newcommand{\baa}{\begin{align}}
\newcommand{\eaa}{\end{align}}
\newcommand{\br}{{\bm r}}

\newcommand{\bk}{{\bm k}}

\newcommand{\eqn}[1]{Eq.~\eqref{#1}}
\newcommand{\fig}[1]{Fig.~\ref{#1}}

\newcommand{\rtab}[1]{Table~\ref{#1}}

\newcommand{\bracket}[3]{\langle #1|#3|#2\rangle}

\newcommand{\ket}[1]{|#1\,\rangle}

\newcommand{\EHxc}{E_{\rm Hxc}}

\newcommand{\inbra}[1]{{#1}}

\newcommand{\etal}{{\it et al.}}

\usepackage{ifthen}
\usepackage{calc}
\usepackage{url}
\usepackage{fancyvrb}
\usepackage[utf8]{inputenc}
\usepackage[english]{babel}%

\usepackage{mathrsfs}

\usepackage{tikz}
\usetikzlibrary{shapes,shadows,arrows}
\usepackage{amsmath}
\let\oldFootnote\footnote
\newcommand\nextToken\relax
\renewcommand\footnote[1]{\oldFootnote{#1}\futurelet\nextToken\isFootnote}
\newcommand\isFootnote{\ifx\footnote\nextToken\textsuperscript{,}\fi}
\makeatletter
\newcommand\footnoteref[1]{\protected@xdef\@thefnmark{\ref{#1}}\@footnotemark}
\makeatother

\begin{document}
\title{First-principles photoemission spectroscopy of DNA and RNA nucleobases 
from Koopmans-compliant functionals}

\author{Ngoc Linh Nguyen} \email{linh.nguyen@epfl.ch} 
\affiliation{Theory and
Simulations of Materials (THEOS), and National Centre for Computational Design
and Discovery of Novel Materials (MARVEL), \'Ecole Polytechnique F\'ed\'erale
de Lausanne, 1015 Lausanne, Switzerland} 
\author{Giovanni Borghi}
\affiliation{Theory and Simulations of Materials (THEOS), and National Centre
for Computational Design and Discovery of Novel Materials (MARVEL), \'Ecole
Polytechnique F\'ed\'erale de Lausanne, 1015 Lausanne, Switzerland}
\affiliation{Centro S3, CNR--Istituto Nanoscienze, 41125 Modena, Italy} 
\author{Andrea Ferretti} 
\affiliation{Centro S3, CNR--Istituto Nanoscienze, 41125 Modena, Italy} 
\author{Nicola \surname{Marzari}}
\affiliation{Theory and Simulations of Materials (THEOS), and National Centre
for Computational Design and Discovery of Novel Materials (MARVEL), \'Ecole
Polytechnique F\'ed\'erale de Lausanne, 1015 Lausanne, Switzerland}
%

\begin{abstract}
The need to interpret ultraviolet photoemission data 
strongly motivates the refinement of first-principles techniques able to accurately predict
spectral properties. In this work we employ Koopmans-compliant functionals, 
constructed to enforce piecewise linearity in approximate density functionals,
to calculate the structural and electronic properties of DNA and RNA nucleobases. 
Our results show that not only ionization potentials and electron affinities are accurately predicted with mean absolute errors $< 0.1$ eV,
but also that calculated photoemission spectra are 
in excellent agreement with experimental ultraviolet photoemission spectra.
In particular, the role and contribution of different tautomers to the photoemission spectra are
highlighted and discussed in detail.
The structural properties of nucleobases are also investigated, showing an improved description 
with respect to local and semilocal density-functional theory.
Methodologically, our results further consolidate the role of Koopmans-compliant functionals in providing, 
through orbital-density-dependent potentials, accurate electronic and spectral properties.
\end{abstract}

\maketitle
\section{Introduction}
The nucleobases adenine (A), cytosine (C), thymine (T), guanine (G), and uracil (U) 
are the primary building blocks of deoxyribonucleic (DNA) and ribonucleic (RNA) acids. 
The sequence of base pairs, which are stacked upon one another leading directly to the helical structure of DNA and RNA, carries all the genetic information of living organisms. 
Due to such biological importance, a proper understanding of the structural and photoelectron properties of these molecules is a priority, in order to unveil the mechanisms of formation of DNA and RNA chain, and the reaction dynamics under exposure to ultraviolet light or ionizing radiation~\cite{colson_structure_1995}. 

Extensive studies to understand the electronic properties of nucleobases have been carried out for many years, both theoretically and experimentally~\cite{hush_ionization_1975, dougherty_photoelectron_1978, choi_ionization_2005, trofimov_photoelectron_2006, schwell_vuv_2008, zaytseva_theoretical_2009,kostko_ionization_2010, roca-sanjuan_ab_2006, roca-sanjuan_ab_2008, qian_photoelectron_2011, foster_nonempirically_2012, JCC:JCC24266}. Notwithstanding such efforts, many questions related to photoelectron properties, such as the stability and symmetry of ionized states, are yet to be understood \cite{trofimov_photoelectron_2006,satzger_reassignment_2006}. 
Furthermore, under experimental conditions, nucleobases appear in several tautomeric or conformeric variants, which differ from one another only in the position of a hydrogen in the structure, and which have energies lying very close to each other. This makes the detailed understanding of spectral properties quite 
challenging, since the contributions of each single isomer are hard to resolve.
Distinguishability can be attained through photoemission experiments with high photon energies, such as X-ray techniques~\cite{feyer_tautomerism_2009, schiedt_anion_1998}; these, however, are not suitable for examining the properties in vivo, and do not address the energies of the valence electrons. For these reasons, 
a synergy of low-energy experiments and accurate theoretical simulations 
would be most beneficial
to interpret photoemission measurements.

Several {\it ab-initio} ground-state calculations have already been carried out in order to determine
which tautomers are the most energetically 
favorable \cite{feyer_tautomerism_2009, bravaya_electronic_2010}. 
Similarly, many efforts have also been devoted to predicting photoemission spectra~\cite{trofimov_photoelectron_2006, zaytseva_theoretical_2009}. These efforts aim at understanding the nature of the spectral peaks,
possibly labeling them with their respective symmetry quantum numbers.

From the theoretical point of view, most photoemission studies have been performed using many-body perturbation theory or high-level wave-function methods (see Ref.~\citenum{qian_photoelectron_2011} and references within), whose considerable computational cost prevents them from being applied to more complex biological environments or to sets of paired bases. 
This is the reason why simpler methods such as Hartree-Fock (HF) or density-functional theory (DFT), 
computationally less demanding, are still frequently employed. Unfortunately, their accuracy in the calculation of ground state energies is not complemented by a comparable precision in predicting electronic excitation energies and photoemission spectra~\cite{Onida2002}. 
These drawbacks are intrinsic to  Kohn-Sham (KS) DFT whose single particle energies 
[except for the highest occupied molecular orbital (HOMO)~\cite{PhysRevLett.49.1691, PhysRevB.56.16021}] cannot 
even in principle be interpreted as quasiparticle excitation energies~\cite{Onida2002,Gatti2007prl, ferr+14prb} 
(though arguments exist suggesting that exact KS eigenvalues may provide good approximations 
to them~\cite{PhysRev.145.561, casida1995pra, chong2002jcp} ).

Recently, Koopmans-compliant (KC) functionals were introduced~\cite{dabo_towards_2008, Dabo2009, Dabo2010, psik_koopmans, dabo_piecewise_2014, Dabo2013, Borghi_PRB_2014} to enforce a generalized criterion of piecewise linearity (PWL) in the energy of approximate DFT functionals with respect to the fractional removal or addition of an electron from any orbital of the system. 
This PWL condition is a generalization to the entire manifold of the molecular DFT+U approach~\cite{kuli+06prl, Kulik_JPC_2008}, stemming from similar linearization ideas in the solid state.~\cite{Cococcioni2005}
The condition of Koopmans' compliance is naturally akin to that of enforcing a correct description of charged excitations~\cite{Dabo2010,nguyen_first-principles_2015}, and can therefore lead to orbital energies that 
are comparable to the quasiparticle excitation energies of photoemission experiments. 
In a previous work~\cite{nguyen_first-principles_2015}, we showed the remarkable performance of KC functionals in predicting ultraviolet photoemission spectra (UPS) and orbital tomography momentum maps for photovoltaic molecules, 
showing an agreement with experiments for frontier orbital energies [ionization potentials (IPs) and electron affinities (EAs)] that is comparable (in some cases even slightly superior) to state-of-the-art methods 
in many-body perturbation theory, while preserving a moderate computational cost and scaling~\cite{Borghi_PRB_2014}, 
and the quality of potential energy surfaces of the underlying DFT functionals~\cite{Borghi_PRB_2014}.

In this work we perform a study on DNA and RNA nucleobases using the best performing KC functional, labeled KIPZ~\cite{Borghi_PRB_2014}, whose performance on small molecules was assessed in Refs.~\citenum{Borghi_PRB_2014} and~\citenum{nguyen_first-principles_2015}. Accuracy in predicting spectroscopic properties of DNA and RNA bases is compared here to experiments, standard DFT calculations, many-body perturbation theory, and quantum-chemistry methods.
We illustrate the effectiveness of the KC approach in distinguishing tautomers and in correctly predicting the geometrical properties of nucleobases (so far accessible 
only via second-order M\o{}ller-Plesset (MP2) perturbation theory~\cite{doi:10.1021/j100063a019}).

The paper is organized as follows: in Sec.~\ref{sec:methodology} we review the formulation of Koopmans-compliant functionals
and provide a simple algorithm to perform self-consistent structural optimizations. 
In the second part (Sec.~\ref{sec:results}) we study the binding energies of the frontier orbitals (Sec.~\ref{sec:binding_energies}) and the photoelectron properties and photoemission spectra (Sec.~\ref{sec:ups}) of the most stable nucleobase tautomers. Finally (Sec.~\ref{sec:geometry_results}), we discuss molecular geometries and show the effects of structural changes on photoelectron properties. 

\section{Methodology}\label{sec:methodology}
The main advantage of KS-DFT over many-body perturbation theory techniques or quantum chemistry methods is that the electronic ground state density of a system is inexpensively parametrized though a set of single-particle orbitals which are the result of the diagonalization of an effective noninteracting KS Hamiltonian. This enables to carry-out calculations that, whilst including many-body effects through the density functional, have the same computational cost of an independent-electron system. All this comes at the expense of having to find reasonable approximations to the exact energy functional, whose exact form remains, excluding some special low-dimensional model systems~\cite{Umrigar_JCP_94}, unknown.
Unfortunately, only the HOMO eigenvalue (negative IP) would be correctly described in exact KS-DFT, but even for this case the most common and computationally inexpensive approximations that are employed provide values that are in poor correspondence with the first particle-removal energies.
Within the local-density approximation (LDA)~\cite{perdew_self-interaction_1981} or the generalized gradient approximations (GGAs) [with a common example as Perdew-Burke-Ernzerhof (PBE) approximation~\cite{perdew_generalized_1996}], 
the HOMO values are systematically too high in energy (underestimating the IP),
in part due to large self-interaction errors (SIE)~\cite{perdew_self-interaction_1981}.
Self-interaction is not only responsible for incorrect electron binding energies, but also for the spatial over-delocalization of charge densities and KS orbital wave functions~\cite{kuli+06prl,cohen_insights_2008}, which in turn affects multiple aspects of ground-state predictions, including molecular equilibrium bond lengths, molecular dissociation energies, adsorption and transition states energies~\cite{vydrov_scaling_down_PZ}.  

\subsubsection{Piecewise linearity and self-interaction errors}

With the purpose of correcting SIE, Perdew and Zunger~\cite{perdew_self-interaction_1981} introduced an orbital-dependent density (ODD) functional, termed here PZ, obtained by subtracting from an approximate functional $E^{\rm app}$ the sum of the Hartree ($E_{\rm H}[\rho_{i}]$) and exchange-correlation ($E_{\rm xc}[\rho_{i}]$) energies of each single (fully or partially) filled orbital $\rho_{i}(\bf r)$:   
\begin{equation}\label{pz_form}
E^{\rm PZ} = E^{\rm app} - \sum_{i}(E_{\rm H}[\rho_{i}] + E_{\rm xc}[\rho_{i}]). 
\end{equation}
While leading to an exact formulation for one-electron systems, the PZ functional generally overcorrects the SIE~\cite{vydrov_scaling_down_PZ} in many-electron systems, resulting incorrect ionization energies, structural predictions, and reaction paths~\cite{vydrov_scaling_down_PZ, Jonsson_JCP_2012}.

As a route towards a good description of the properties of many-electron systems, several authors~\cite{Cococcioni2005, kuli+06prl, cohen_insights_2008, kraisler_piecewise_2013} have suggested a new definition of the SIE based on 
the lack of PWL of the total energy as a function of the (fractional) number of electrons.
Indeed, it can be shown that a dependence of $E(N)$ on the particle number $N$ that is convex tends to delocalize total and orbital densities, whereas functionals for which $E(N)$ is concave (such as PZ) lead to over-localization~\cite{vydrov_tests_2007}. 

As mentioned, KC functionals~\cite{dabo_towards_2008, Dabo2009, Dabo2010, psik_koopmans, dabo_piecewise_2014, Dabo2013, Borghi_PRB_2014} can be seen as a generalization of DFT+U aimed at explicitly enforcing PWL to an entire electronic manifold. 
These functionals are obtained by removing, orbital-by-orbital, the non-linear (Slater) contribution 
to the total energy and by replacing it by a linear (Koopmans) term. This linear term is chosen either following Slater's suggestion~\cite{Slater1974}, i.e., proportional to the orbital energy at half orbital filling 
(in which case the KC functional is simply labeled K~\cite{Dabo2010, Borghi_PRB_2014}), or as the difference between the energies of the two adjacent electronic configurations with integer occupation; this latter is labeled KI 
(``I'' standing for ``integral''). The numerical differences between these two approximations are largely negligible, 
and we focus here on KI, which is simpler to implement. 
KI functionals, described in detail in
Ref.~\citenum{Borghi_PRB_2014}, are obtained from an 
approximate functional $E^{\inbra{{\rm app}}}$ as 
\begin{equation}\label{Eq:en_KI}
E^{\inbra{{\rm KI}}}=E^{\inbra{{\rm app}}} + \sum_{i} \alpha_{i} \Pi^{\rm KI}_i\,, 
\end{equation}
where
\begin{align}\label{Eq:corr_KI} 
\Pi^{\inbra{\rm KI}}_{i} &= -\EHxc[\rho]+\EHxc[\rho-\rho_{i}] \nonumber\\ &+f_{i}\robra{-\EHxc[\rho-\rho_{i}]+\EHxc[\rho-\rho_{i}+ n_{i}]}\,.
\end{align}
In the above equations, $E_{\rm Hxc} = E_{\rm H} + E_{\rm xc}$, $n_{i}(\br) = |\varphi_{i}(\br)|^2$, 
$\rho_{i}(\br) = f_{i} |\varphi_{i}(\br)|^2$, $\rho(\br) = \sum_{i} {f_{i} |\varphi_{i}(\br)|^2}$, and $\alpha_i$ are orbital-dependent screening coefficients that account for orbital relaxations, since for $\alpha_{i}=1$, the KI functional described in \eqn{Eq:en_KI} fulfills exactly the generalized Koopmans condition at frozen orbitals~\cite{psik_koopmans}. The $\alpha_i$ coefficients, which can be computed from first principles (see below), are generally smaller than one. 
It should be noted that the KI functional is piecewise linear with respect to fractional changes in the particle number, 
but it does not change the total energy nor the ground-state wave function
(and consequently the one-body density matrix) of the approximate "base" functional whenever the
system has an integer number of particles.
We also note in passing that, using this
terminology, the ensemble-DFT correction of Ref.~\citenum{kraisler_piecewise_2013}
is equivalent to the KI functional, when applied to the frontier orbitals.

It is possible to modify the definition in~\eqn{Eq:en_KI} so as to obtain a functional which is, similarly to PZ, exact in the one-electron limit and variational, while remaining approximately self-interaction free in the many-electron case (we stress that the generalized Koopmans condition is stronger than just being many-electron self-interaction free, since it applies simultaneously to all orbitals). 
Such functional can be obtained by applying a KI correction on top of the PZ functional, resulting in the following definition:
\begin{equation}
\label{eq:en_KIPZ}
E^{\rm KIPZ} = E^{\rm app} + \sum_{i} \alpha_i \cubra{{\Pi^{\rm KI}_i - f_{i} E_{\rm Hxc} [n_{i}]}}.
\end{equation}
The Koopmans orbital-by-orbital linearity condition imposed through Eq.~\eqref{Eq:en_KI} or~\eqref{eq:en_KIPZ} leads to an ODD formulation in which the energy functional depends on the density of the individual orbitals.
As such, differently from DFT functionals, but similarly to other ODD functionals such as PZ, KC functionals are not invariant under unitary transformations within the manifold of filled orbitals
~\cite{Borghi_PRB_2014, hofmann_using_2012, lehtola_variational_2014, ferr+14prb, Borghi_PRB_2015}
and the {\it variational orbitals} $\ket{\varphi_i}$ that minimize the functional are different from the eigenstates or {\it canonical orbitals} $\ket{\phi_m}$ that diagonalize the matrix of Lagrange multipliers, as discussed, e.g., in Refs.~\citenum{hofmann_using_2012,lehtola_variational_2014,Borghi_PRB_2015}.
The strategy that we use to minimize KC functionals, which follows the ensemble-DFT algorithm~\cite{marzari_ensemble_1997} (note that this is unrelated to the ensemble-DFT correction of Ref.~\citenum{kraisler_piecewise_2013}) for the case of orbital-density-dependent
functionals, consists of two nested steps: (i) a minimization with respect to unitary
transformations at fixed orbital manifold (inner loop), that leads to a projected, 
unitary-covariant functional of the orbitals only enforcing the Pederson condition~\cite{pederson-jcp-1984}; 
(ii) a variational optimization of the orbital manifold of this
projected functional~\cite{Borghi_PRB_2015} (outer loop).

The generalized eigenvalue equation obtained within the ODD formalism reads
\begin{equation}
\label{Eq:KC_h}
\hat{H}^{\rm app}\ket{\phi_{m}} + \hat{\Sigma}_{m} \ket{\phi_{m}} =  \varepsilon_{m} \ket{\phi_{m}}
\end{equation}
with 
\begin{equation}\label{eq:KIPZ_eigenvalue_equation}
\Sigma_{m}(\br) \phi_{m}(\br) = \sum_{i} v^{\text{ODD}}_{i}(\br) \varphi_{i}(\br) U^\dagger_{im}
\end{equation}
where the unitary matrix $U$ transforms the variational orbitals $\ket{\varphi_{i}}$ 
into the canonical ones $\ket{\phi_{m}}=\sum_{i} \ket{\varphi_{i}} U^\dagger_{im} $,
and the ODD potential is given by
\begin{equation}
\label{Eq:pot_KI}
{v}^{\text{ODD}}_{i}(\br) = \sum_j \frac{\delta \Pi^{\rm ODD}_j}{\delta \rho_{i}(\br)}.
\end{equation}
Full details about the expressions for ${v}^{\text{ODD}}_{i}(\br)$ when the ODD correction is KI or KIPZ
are given in Ref.~\citenum{Borghi_PRB_2014}.
As discussed in Refs.~\citenum{ferr+14prb} and~\citenum{nguyen_first-principles_2015}, ODD canonical 
orbitals can be interpreted as approximations to Dyson orbitals (solutions of quasiparticle equations), and their energies as particle-removal energies; this interpretation will be supported by the results of the present work.

\subsubsection{Calculation of the screening coefficient}

In the definitions of KI and KIPZ functionals, the multiplicative factors $\alpha_i$ are 
meant to account for orbital relaxation~\cite{psik_koopmans} and should be orbital dependent. 
Often, we simply choose them to be all equal to a single (effective) $\alpha$ value chosen so that the IP
of a neutral molecule is equal to EA of the molecular cation~\cite{Dabo2010, PhysRevLett.105.266802}.
The optimal value of $\alpha$ is computed through the secant recursion method:
\begin{equation}\label{eq:alpha_calc_0}
 \alpha_{n+1} = \alpha_n + \frac{(1-\alpha_n)(\epsilon^{\alpha_n}_{L, N-1} - \epsilon^{\alpha_n}_{H, N})}{(\epsilon^{\alpha_n}_{L, N-1} - \epsilon^{\alpha_n}_{H, N}) - (\epsilon^{\alpha_{n-1}}_{L, N-1} - \epsilon^{\alpha_{n-1}}_{H, N}) }
\end{equation}
where $\epsilon^{\alpha_n}_{L, N-1}$ and $\epsilon^{\alpha_n}_{H, N}$ are the LUMO energy of the positively ionized system and the HOMO energy of the neutral system, respectively. As already mentioned, the procedure to determine $\alpha$ is fully {\it ab initio}, since all HOMO and LUMO eigenvalues appearing in \eqn{eq:alpha_calc_0} are the result of numerical simulations involving an approximate DFT functional with Koopmans' corrections. 
As shown by Dabo et al.~\cite{Dabo2010} and Borghi et al.~\cite{Borghi_PRB_2014}, a constant screening coefficient is sufficient to accurately predict IPs and EAs of a variety of atomic and molecular systems, even though an orbital-dependent $\alpha$ might be more accurate in the case of large or extended systems, or in systems composed of sub-systems having very different orbital relaxation properties. 

In fact, when computing EAs of molecules, results improve~\cite{nguyen_first-principles_2015} when choosing as screening coefficient the one appropriate for the LUMO, which can be obtained easily from \eqn{eq:alpha_calc_0} replacing $N$ with $N+1$. This ``anionic'' screening, which we will refer to as $\alpha^{(c)}$, where the letter ``c'' stands for ``conduction'' (as opposed to the ``neutral'' or ``valence'' one, for which we will use the symbol $\alpha^{(v)}$) leads to much better results for the binding energies of empty orbitals. 
In Sec.~\ref{sec:binding_energies} and \rtab{tab_vip_ea_all}, we will show that this choice reproduces correctly the energies and the orbital ordering of the empty states of nucleobases. Meanwhile, we show in~\rtab{tab_values_alpha} the values of $\alpha^{(v)}$ and $\alpha^{(c)}$ computed from \eqn{eq:alpha_calc_0} for the most stable nucleobases. One can immediately 
remark that $\alpha^{(c)}$ is very close to 1 (except for the case of the T molecule which has $\alpha^{(c)} > 1$, meaning that the LUMO state is more delocalized compared to the other orbitals), suggesting that the orbital relaxation that takes place in the anionic system upon removal of its extra electron is very small.

\subsubsection{Geometry and screening optimization}\label{sec:screen_opt}

\begin{figure}
  \includegraphics[width=0.45\textwidth]{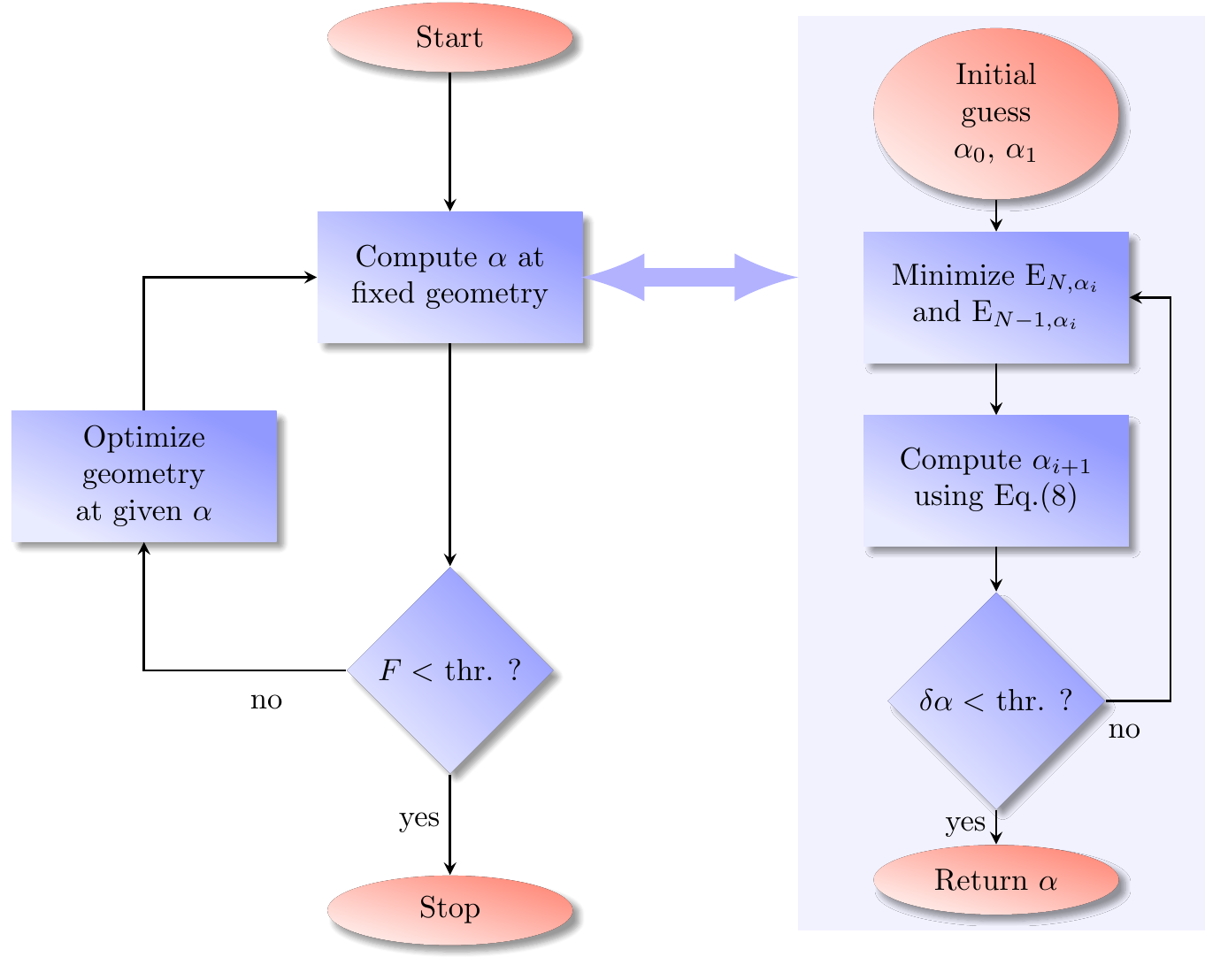}
  \caption{\label{fig:scf_flow} Diagram of the self-consistent KIPZ scheme used to optimize the screening factor $\alpha$ ($\alpha^{(v)}$) along with molecular structure.} 
\end{figure}

In previous work~\cite{Dabo2010, psik_koopmans,Dabo2013, Borghi_PRB_2014, nguyen_first-principles_2015} the calculation of $\alpha^{(v)}$ was performed at fixed geometry, assuming the 
variation of the screening during the geometry optimization to be negligible.
In this work we explore self-consistency for $\alpha^{(v)}$ with respect to molecular geometry. The workflow is presented in Fig.~\ref{fig:scf_flow}, and starts with an initial guess that can be chosen from the geometries computed from standard DFT. 
Then, the value $\alpha^{(v)}$ corresponding to this geometry is estimated according to \eqn{eq:alpha_calc_0}.

After that, a geometry optimization is performed with the KIPZ functional, using the computed $\alpha^{(v)}$ as screening coefficient. A new value of $\alpha^{(v)}$ is then calculated for the new geometry. This completes a self-consistent loop which is iterated until the inter-atomic force ($F$) and the change in screening coefficient ($\Delta \alpha$) between subsequent steps are smaller than a given threshold. In Table~\ref{tab_values_alpha}, we show
the change of the $\alpha$-coefficients during the self-consistent procedure. The detailed results obtained with this optimization scheme are reported in Sec.~\ref{sec:geometry_results}, where we discuss the accuracy
of Koopmans-compliant functionals in predicting the geometry of DNA and RNA bases.
%
\begin{table}
\caption{\label{tab_values_alpha} Values of $\alpha^{(v)}$ and $\alpha^{(c)}$ computed on top of the initial structures, PBE (@PBE), or on top of the structure resulting from scf-KIPZ optimization (@KIPZ) of different DNA/RNA nucleobases.}
\begin{ruledtabular}
\begin{tabular}{l c c c c }
    & $\alpha^{(v)}$@PBE& $\alpha^{(v)}$@KIPZ & $\alpha^{(c)}$@PBE & $\alpha^{(c)}$@KIPZ \\
\hline
A &0.4523 & 0.4713&0.9822 &0.9934 \\
T &0.4654 & 0.4732&1.0324 &1.0184 \\
U &0.4801 &0.4886 &0.9422 &0.9681\\
C$_1$&0.4389 &0.4560&0.9725 & 0.9873\\
G$_2$&0.4225 &0.4465&0.9352 & 0.9523 \\
\end{tabular} 
\end{ruledtabular}
\end {table}

\section{Results and Discussion}\label{sec:results}
\subsection{Technical details}\label{sec:technical_details}
In this Section we present the results obtained by applying the above scf-KIPZ procedure to compute binding energies, ultraviolet photoemission spectra, and optimized geometries for all nucleobases. 
Since our code (a modified version of cp.x from the Quantum-ESPRESSO distribution~\cite{giannozzi2009jpcm}) 
works with periodic-boundary conditions, we place the molecules inside an orthorhombic cell 
with at least 18 Bohr of vacuum on each side, and we add 
reciprocal-space counter-charge corrections~\cite{li-dabo11prb} 
to the electrostatic energies and potentials 
in order to suppress the Coulomb interactions between periodic replicas. 
All calculations are performed using norm-conserving pseudopotentials~\cite{PP} and the cutoff for the 
plane wave expansion of wavefunctions is set to 60 Ry. Moreover, in all PZ and KIPZ calculations presented in this paper, the orbital-density dependent corrections (either PZ or KIPZ) are computed on top of the PBE functional. 
In previous work, we showed that the combination of PZ and KIPZ with PBE as base functional produces the best results for the electronic eigenvalue spectrum when minimized on the Hilbert space of complex wave functions~\cite{klupfel_importance_2011, Borghi_PRB_2014}, which is the procedure adopted here.

\begin{figure}
  \includegraphics[scale=0.32]{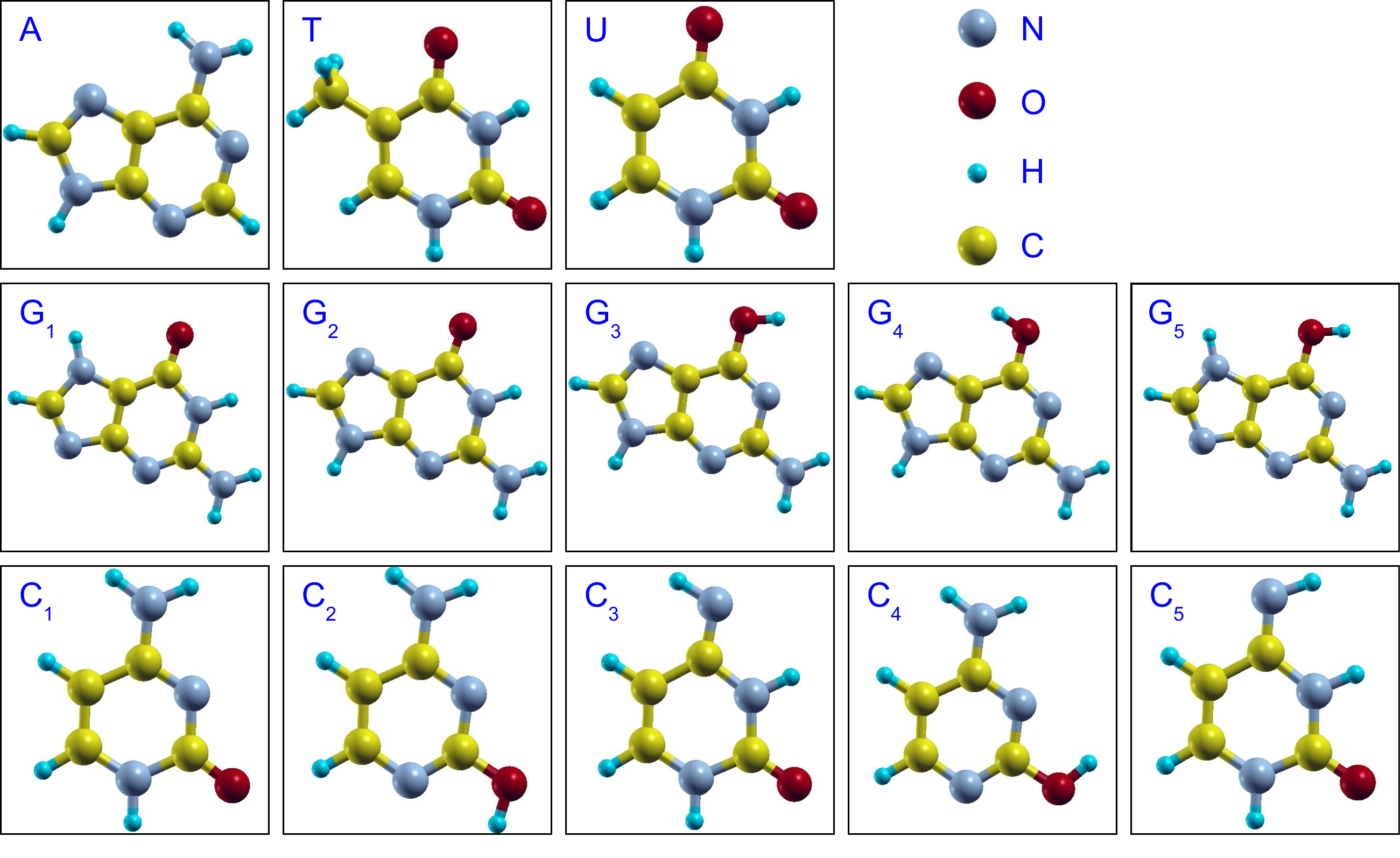}
  \caption{\label{fig:struc_properties} Atomic structure of A, T, U, and of the five most stable tautomers of G and C considered in this study.} 
\end{figure}

The set of all possible nucleobase tautomers that could be the subject of our study is quite large; in this work we choose to study only the most stable ones, as reported in~\fig{fig:struc_properties}. From our DFT calculations we find that there is a single tautomer that is energetically favored for A, T, and U (for these bases the total energy difference between the less stable tautomers and the most stable one is $> 0.43$ eV. This fact yields the normalized Boltzmann weighting factors evaluated at room temperature (300 K), computed using the total energies and neglecting vibrational entropy, for the most stable tautomers are one and the others are zero). Instead, for the C and G nucleobases, we find five different tautomeric forms with very close total energies, and we include all of them in our set. For each nucleobase, we compute the KIPZ total energy of different tautomers using only one value of $\alpha$, which, for C and G, is computed as the average of the screening coefficients of the 5 tautomers. 
The normalized Boltzmann weighting factors for the five tautomers of C and G are C1 :C2 :C3 :C4 :C5 =0.32:0.47:0.12:0.06:0.03 and G1 :G2 :G3 :G4 :G5 =0.44:0.20:0.18: 0.16 : 0.02, respectively
.
These ratios are in reasonable agreement (correct order) with coupled-cluster CCSD(T) 
results\cite{B202156K,plekan_experimental_2009},
giving ${\rm C}_1:{\rm C}_2:{\rm C}_3 = 0.23:0.64:0.13$ and 
${\rm G}_1:{\rm G}_2:{\rm G}_3=0.60:0.27:0.13$ .
In Sec.~\ref{sec:ups} we show how the experimental UPS of C and G can be reproduced with good accuracy from the average of the spectra of different tautomers using the KIPZ weighting factors.
\subsection{Binding energies of frontier orbitals}\label{sec:binding_energies}

In this section we assess the accuracy of the KIPZ functional in computing vertical IPs and EAs, and more in general in predicting the binding energies and the molecular orbital (MO) character of the frontier occupied and unoccupied orbitals. 

\begin{table*}
\caption{\label{tab_vip_ea_all} Binding energies and molecular orbital characters of HOMO-1 ($\sigma$), HOMO ($\pi$), LUMO (dipole-bound [DB] state) and LUMO+1 (valence-bound [VB] state) canonical orbitals obtained from KIPZ eigenvalues in comparison with  startdard DFT (PBE exchange-correlation functional), PZ, G$_0$W$_0$, scf-GW and quantum chemistry
calculations (CASPT2: complete active space with second-order perturbation theory; CCSD(T): coupled-cluster with
singles, doubles, and perturbative triple excitations) and experimental data. Experimental values are taken as a reference for the calculation of mean absolute errors (MAE). Orbital energies within KIPZ are obtained with the $\alpha^{(v)}$ screening factor for ``valence" filled orbitals, and with $\alpha^{(c)}$ for empty ``conduction" orbitals. For the empty orbitals we report, in parentheses, orbital energies obtained using the $\alpha^{(v)}$ screening factor instead.}

\begin{ruledtabular}
\begin{tabular}{crrrrrrrr}
   &KIPZ orb.& PBE&PZ&G$_0$W$_0$(PBE)\footnote{\label{qian}Reference~\onlinecite{qian_photoelectron_2011}}&scf-GW\footnote{\label{faber}Reference~\onlinecite{faber_first-principles_2011}}&CASPT2/CCSD(T)\footnote{\label{roca1}Reference~\onlinecite{roca-sanjuan_ab_2008}}\footnote{\label{roca2}Reference~\onlinecite{roca-sanjuan_ab_2006}}\footnote{\label{maciej}Reference~\onlinecite{haranczyk_valence_2005}} &	KIPZ &Experiment\footnote{\label{expt1}Collected in reference~\onlinecite{roca-sanjuan_ab_2006}}\footnote{\label{expt2}Collected in reference~\onlinecite{roca-sanjuan_ab_2008}}\footnote{\label{expt3}Reference~\onlinecite{trofimov_photoelectron_2006}}\footnote{\label{expt4}Reference~\onlinecite{Dougherty1978379}}\footnote{\label{expt5}Reference~\onlinecite{zaytseva_theoretical_2009}}\footnote{\label{expt6}Reference~\onlinecite{schiedt_anion_1998}}\footnote{\label{expt7}Reference~\onlinecite{hendricks_dipole_1996}} \footnote{\label{expt8}Reference~\onlinecite{desfrancois_electron_1996}}\\
\hline					
A&[L+1]VB &1.67&1.76	&-0.25&	-1.14& -0.91\footnoteref{roca1} &(-0.32)-0.47 &-0.56$\sim$-0.45\footnoteref{expt2}\\
 & [L]DB & 0.72 &0.63 & -0.31 & &  & (0.13)-0.02& 0.012 \footnoteref{expt8}\\
 &[H]$\pi$&5.55&9.90&7.99&	8.22& 8.37\footnoteref{roca2}/8.40\footnoteref{roca2}&8.41&8.3$\sim$ 8.5\footnoteref{expt1}/8.47\footnoteref{expt3} \\
 &[H-1]$\sigma$ &5.82& 10.81	&	8.80&	9.47& 9.05\footnoteref{roca2}&9.01 &9.45\footnoteref{expt3}\\
\hline						 
T&[L+1]VB &2.29&2.49  &0.24&-0.67& -0.60\footnoteref{roca1}/-0.65\footnoteref{roca1}&(0.18)-0.32 & -0.53 $\sim$ -0.29\footnoteref{expt2} \\
 & [L]DB    &0.59&0.92     & -0.26 &	&  &(0.33)0.06 & 0.062 $\sim$ 0.068\footnoteref{expt7}\\
&[H]$\pi$       &    6.08&10.92 &8.63&9.05&  9.07\footnoteref{roca2}/9.04\footnoteref{roca2} &9.02 &9.0 $\sim$ 9.2\footnoteref{expt1}/9.19\footnoteref{expt3}\\
&[H-1]$\sigma$ 	      &6.11&12.25 &8.94&10.41 &9.81\footnoteref{roca2}&9.77 &9.95 $\sim$ 10.05\footnoteref{expt1}/10.14\footnoteref{expt3} \\
\hline						
U&[L+1]VB&2.44&	2.64& 0.23&-0.64 & -0.61\footnoteref{roca1}/-0.64\footnoteref{roca1} &(0.26)-0.36 & -0.3$\sim$-0.22\footnoteref{expt2}\\
 &[L]DB  &0.82&     0.98         &-0.29&	 &                                                   &(0.38)0.08 & 0.093\footnoteref{expt7} \\
&[H]$\pi$&6.17&	11.22&	8.99&9.47& 9.42\footnoteref{roca2}/9.43\footnoteref{roca2}&	9.45& 9.4$\sim$9.6\footnoteref{expt1}\\
 &[H-1]$\sigma$&6.4&	12.36&		9.07&	10.54& 9.83\footnoteref{roca2} &	9.97 & 10.02 $\sim$ 10.13\footnoteref{expt1}\\
\hline						 
G$_2$&[L+1]VB&1.29&1.53 &-0.43&-1.58& -1.14\footnoteref{roca1} &(-0.17)-0.36 \\
 & [L]DB  &0.63 &0.67 &  &-0.20	& 0.056 $\sim$ 0.065\footnoteref{maciej} &(0.46)0.08 \\
&[H]$\pi$	&5.22&9.62&7.64&7.81& 8.09\footnoteref{roca2}/8.09\footnoteref{roca2} &8.07& 8.0  $\sim$ 8.30\footnoteref{expt1}/8.30\footnoteref{expt4}/8.26\footnoteref{expt5}\\
&[H-1]$\sigma$&5.86&11.56&8.67&9.82&  9.56\footnoteref{roca2}&9.25 & 9.90\footnoteref{expt4}/9.81\footnoteref{expt5}\\
\hline	
C$_1$&[L+1]VB&2.01&2.19&-0.02&-0.91& -0.69\footnoteref{roca1}/-0.79\footnoteref{roca1}&(-0.03)-0.41&-0.55$\sim$-0.32\footnoteref{expt2}\\
 &[L]DB &0.62 & 0.90& -0.23 &	&  &(0.38)0.11& 0.23\footnoteref{expt6} \\
&[H]$\pi$	&5.70&10.50&8.18&8.73& 8.73\footnoteref{roca2}/8.76\footnoteref{roca2}&8.70&8.80$\sim$8.90\footnoteref{expt1}/8.89\footnoteref{expt3}\\
&[H-1]$\sigma$	&6.26&10.96&	8.5&9.89& 9.42\footnoteref{roca2}&9.12&9.45\footnoteref{expt4}/9.55\footnoteref{expt3}\\
\hline
MAE&[L+1]VB&2.51& 2.67& 0.45 & 0.44& 0.32 & (0.43)0.06 \\
 &[L]DB &0.59 &0.76 & 0.29 &	&  & (0.21)0.05 \\	
 &[H]$\pi$ &3.08 &1.61 & 0.54 &0.17& 0.08  & 0.09 \\
&[H-1]$\sigma$ &3.70 &1.80 &0.99 &0.25& 0.25  & 0.36  \\		
\end{tabular} 
\end{ruledtabular}
\end{table*}

\begin{figure}
  \includegraphics[width=0.45\textwidth]{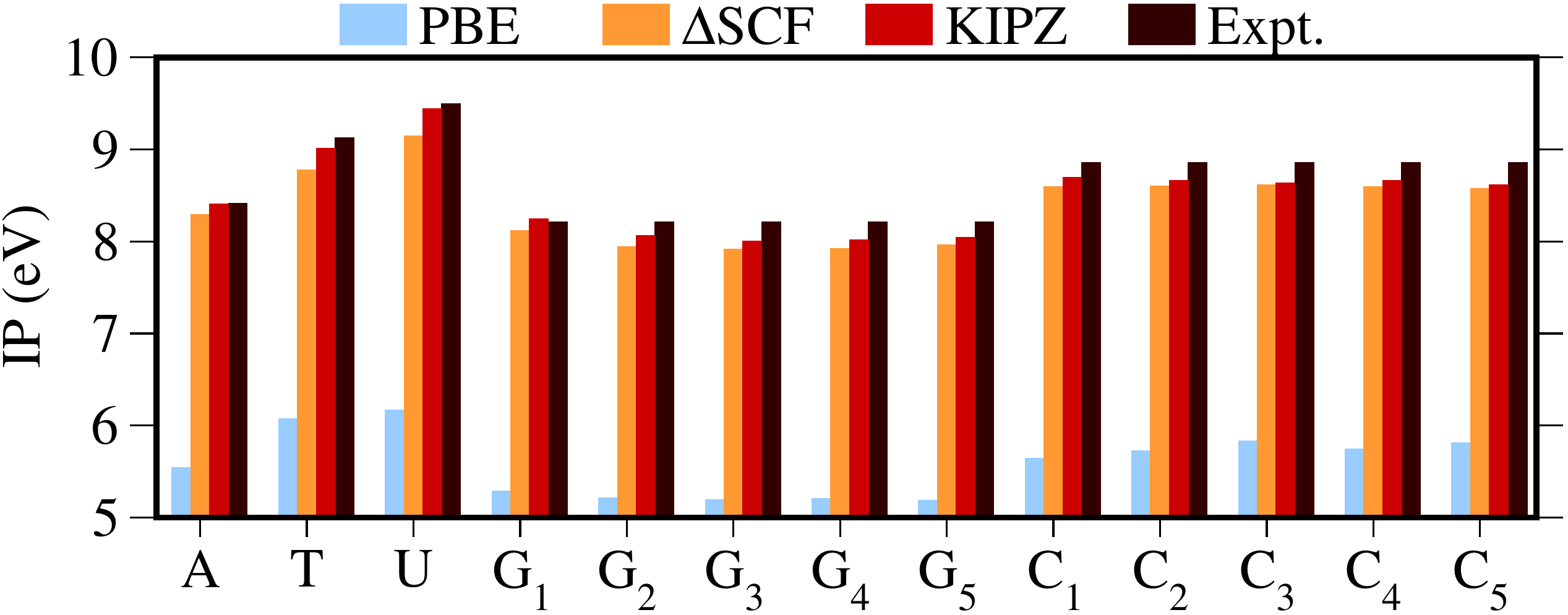}
  \caption{\label{ip_ea_alls} Vertical IPs computed with the standard DFT (PBE exchange-correlation functional), $\Delta$SCF, and KIPZ, for all the nucleobases and their tautomers considered in this work. The experimental numbers used as a reference for vertical ionization potentials are the average values computed from the various experiments listed in~\rtab{tab_vip_ea_all}.} 
\end{figure}

\begin{figure}
\begin{center}
  \includegraphics[scale=0.5]{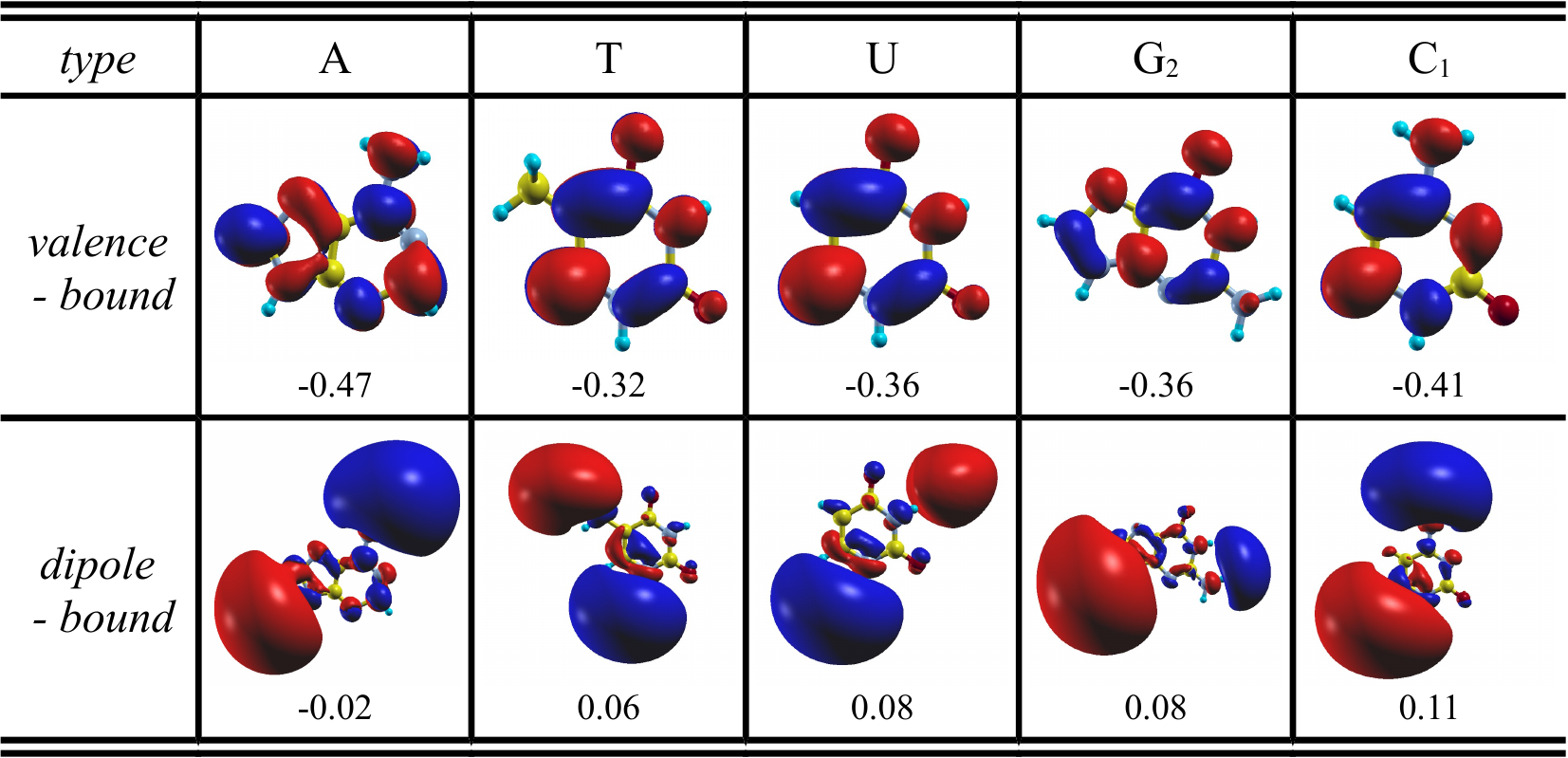}
  \vspace{0.3cm}
  \includegraphics[scale=0.32]{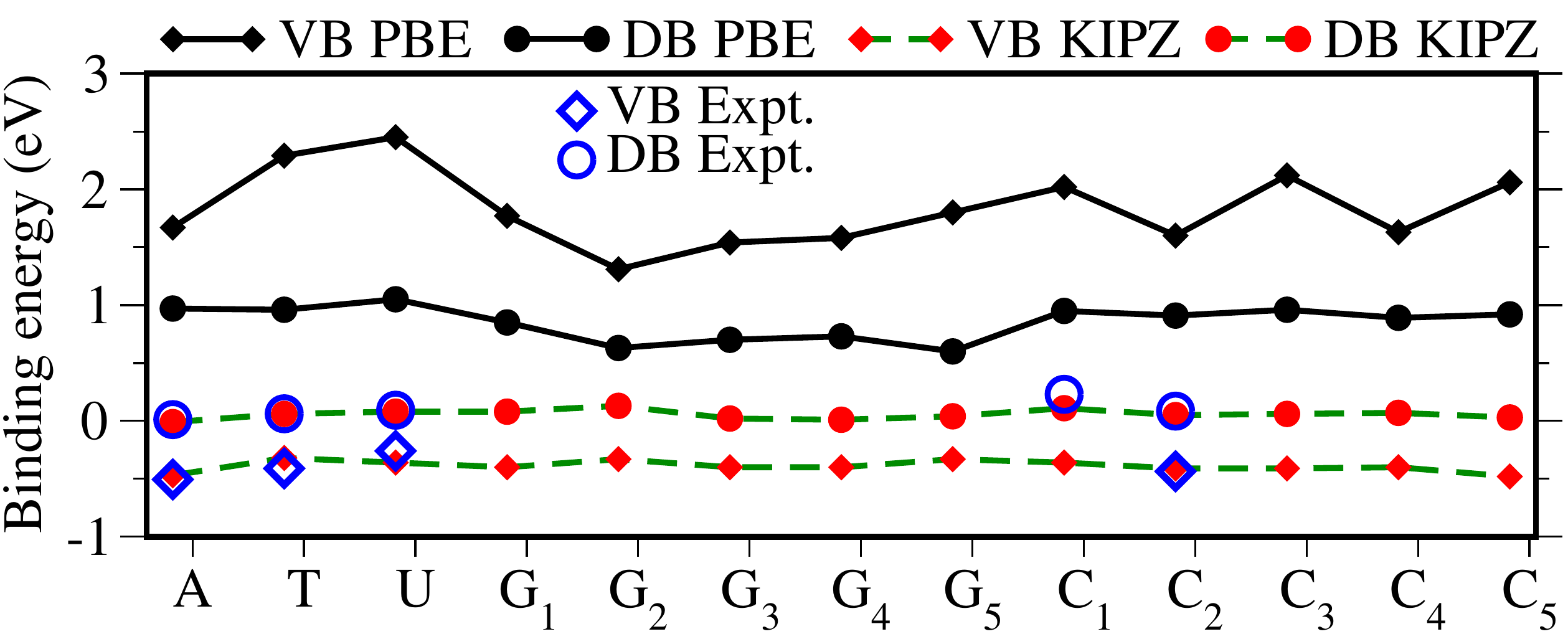}
  \caption{\label{iso_plot_lumo} (Top panel) density isosurfaces for valence-bound (VB) and dipole-bound (DB) unoccupied states of the five DNA nucleobases shown in~\rtab{tab_vip_ea_all}. All are computed with the KIPZ functional. The two different colors refer to the sign of the wave functions. The values listed below the plots are the corresponding orbital binding energies.
  (Bottom panel) Orbital binding energies of VB and DB orbitals computed with the standard DFT (PBE exchange-correlation functional) and KIPZ for the nucleobases considered in this work. The experimental numbers used as a reference are taken from the experimental EA energies listed in~\rtab{tab_vip_ea_all}.} 
\end{center}
\end{figure}

\begin{figure}
  \includegraphics[scale=0.3]{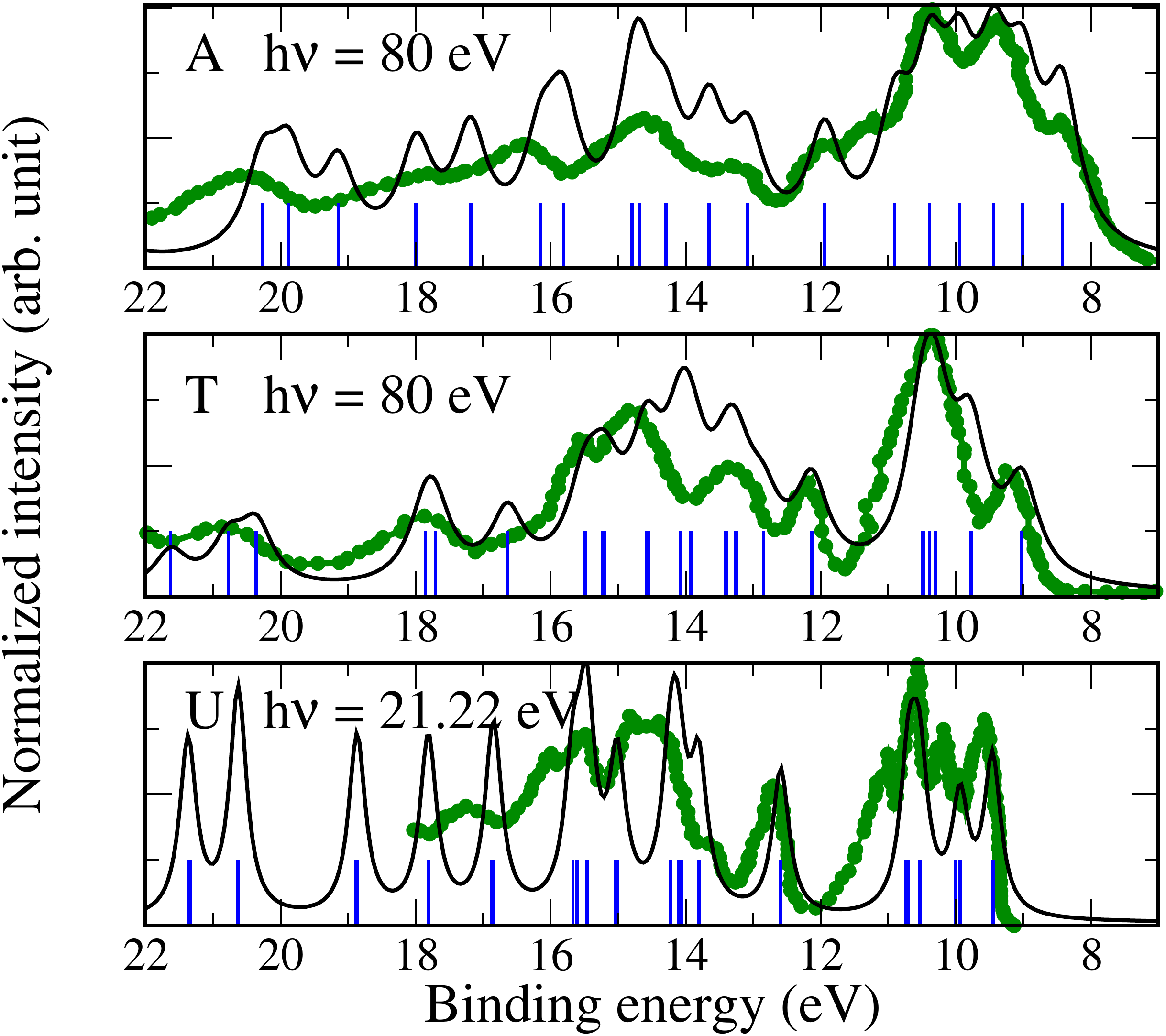}
  \caption{\label{ups_atu} UPS spectra for adenine, thymine and uracil molecules calculated using the KIPZ functional (black line), and plotted as a function of electron binding energy ($h\nu-\hbar^2 k^2/(2m)$). For adenine and thymine the calculations are carried out for an incoming photon energy of 80 eV, and compared with experimental gas-phase UPS measurements (green dots) using radiation with the same photon energy~\cite{trofimov_photoelectron_2006}. For uracil both experimental and simulation data are for a 21.22 eV~\cite{odonnell_ab_1980} photon energy. The blue bars in each plot mark the energies of KIPZ canonical orbitals.} 
\end{figure}

In Fig.~\ref{ip_ea_alls} we present the KIPZ predictions for the first-IP of DNA and RNA bases as compared 
to experiments, together with results computed at the DFT-PBE level and using the Slater $\Delta$SCF method. 
The latter is often considered an accurate approach for computing the first IP and EA for small, finite systems. Experimental data resolved for each C and G tautomer cannot be obtained, due to the closeness in energy of the tautomers (see the discussion at the end of Sec.~\ref{sec:technical_details}); we therefore use the same value as a reference for all of them.

From the data in Fig.~\ref{ip_ea_alls} we find that within PBE the value of the IP is underestimated with a mean absolute error (MAE) of about 3.10 eV with respect to experiments. The incorrect predictions are cured by KIPZ, which reduces the error down to 0.1 eV. The accuracy of KIPZ compares favorably to that of the Slater $\Delta$SCF method, which has a MAE of $\sim$0.15 eV. We also recall here that all the theoretical predictions for PBE-, PZ-, GW- and KIPZ- IPs are obtained from the negative of the orbital energy of the HOMO.

In Table~\ref{tab_vip_ea_all} we show a comparison between KIPZ predictions for frontier orbital binding energies (i.e. HOMO-1, HOMO, LUMO and LUMO+1) and those of other theoretical approaches, including experimental values when available.
The accuracy of KIPZ is found higher than that of PZ, G$_0$W$_0$, and self-consistent GW (scf-GW). 
We note that KIPZ compares favorably also with highly-accurate quantum chemistry methods such as 
complete active space with second-order perturbation theory (CASPT2) and coupled-cluster with
singles, doubles, and perturbative triple excitations [CCSD(T)]. In Table~\ref{tab_vip_ea_all} we report also the binding energies of the first two empty orbitals, recalling again that we take the binding energy of the LUMO orbital as its vertical EA.

Besides the experimental difficulties, the theoretical calculation of nucleobase EAs is also non-trivial, 
mostly because of the large polarity of DNA bases (dipoles larger than 2.5 Debye). Such polarity allows for the existence of a stable, but very weakly bound, ``dipole-bound'' (DB) anionic state. This state is close in energy to an anionic valence bound (VB) state, characterized by an extra electron occupying a valence anti-bonding molecular orbital. 
The nature of these two different states can be investigated using different experimental techniques. The energy of DB states, which are very weakly bound but stable, can be measured from negative ion photoelectron spectroscopy~\cite{hendricks_dipole_1996}. 
VB states instead are accessible through electron transmission spectroscopy~\cite{aflatooni_electron_1998}.
 With this technique it was proved that adding an electron to a VB state requires a positive energy, which means that the electronic configuration of a nucleobase with an extra electron on its anti-bonding orbital is unstable. 
When it comes to numerical simulations, the instability of a state can be established from the negative sign of its binding energy.
Interestingly, not all theoretical methods are able to access both types (VB and DB) of anionic states. Accurate CCSD(T) simulations~\cite{roca-sanjuan_ab_2008} on neutral and negatively charged DNA nucleobases, for instance, result in negative EAs, and predict the anionic state to be of the VB type. This disagrees with experimental findings which suggest the existence of a weakly-bound DB state.
The inability of these CCSD(T) calculations to predict the existence of a frontier DB state has been questioned~\cite{roca-sanjuan_ab_2008} and connected to the basis sets used in the simulations, unable to describe orbitals with a highly diffuse character in the vacuum region around the molecule. 
The same issue affects GW calculations when they are performed with localized basis sets~\cite{faber_first-principles_2011}, while it does not apply to plane-wave-based GW methods. Indeed Qian, Umari and Marzari~\cite{qian_photoelectron_2011} were able to obtain an empty VB state with negative energy (positive binding energy, thus able to bind an extra electron) in the spectrum of neutral guanine.

Using our plane-wave based code we are able to reproduce both VB and DB states, and their orbital densities and energies are shown in~\fig{iso_plot_lumo}. By performing PBE and KIPZ calculations on neutral nucleobases, we can access the orbital energies of empty states and assess their ability to bind extra electrons to the molecule.
PBE and KIPZ orbital energies for VB and DB empty states are shown in~\fig{iso_plot_lumo}. PBE results not only deviate dramatically from experiment, but reverse the order of VB and DB energies, and predict both states to be able to bind electrons. 
The addition of Koopmans' corrections reverses the order of orbital energies and pushes the binding energies of VB states to negative values, restoring the agreement with experimental results~\cite{aflatooni_electron_1998}, showing VB states to be unstable. The agreement of our results for guanine with the G$_0$W$_0$ results of Qian~{\etal}~\cite{qian_photoelectron_2011} is a further proof of the reliability of the KIPZ functional. In~\rtab{tab_vip_ea_all} we show the binding energies of frontier orbitals compared to experiments and different theoretical results. It is worth to stress again that for an accurate prediction of the binding energies of empty orbitals we use as screening coefficient not the screening coefficient of the neutral system ($\alpha^{(v)}$), but the one 
of the anionic system ($\alpha^{(c)}$).

\subsection{Ultraviolet photoemission spectroscopy}\label{sec:ups}

The capability of the KIPZ functional of predicting not only the binding energies of frontier orbitals, but also those of deeper states, makes it a promising tool for the calculation of photoemission spectra. In this section we show how KIPZ 
can describe the position of (quasiparticle) photoemission peaks as well as their strengths and shapes. We obtain theoretical photoemission spectra following the well established three-step model within the sudden approximation~\cite{RevModPhys.75.473}. 
This approach treats the photoexcitation as a transition from an electronic initial state $\ket{\Phi^{\rm N}_{0}}$ --- which is the ground state with energy $E^{\rm N}_{\rm 0}$ --- into an excited $\rm N$-particle state $\ket{\Phi^{\rm N}_{i,\bk}}=\ket{\Phi^{\rm N-1}_{i}}\otimes \ket{\xi_{\bk}}$ of energy $E^{\rm N}_{i,\bk}$, built from the $i^{\rm th}$ excited state of the singly ionized system (with energy $E^{\rm N-1}_{i}$) and the wave function $\xi_{\bk}$ of the ejected electron, approximated by a plane wave with wave vector $\bk$. Here, it is worth stressing that $\xi_{\bk}$ can be approximated in different ways, the simplest one being a plane wave, which can be further orthogonalized to the initial states (with the aim of improving the description of the final states). However, the discrepancy between these approaches has been shown to emerge mainly when looking at states with large binding energies~\cite{Elison_jcp_1974} (low kinetic energy photoelectrons), so in the present work we 
adopt the plane wave approximation, which is expected to be accurate for high kinetic energy photoelectrons. 
The total photoemission intensity can be described, to first order in perturbation theory, through the Fermi's golden rule~\cite{RevModPhys.75.473} as
\begin{equation}\label{pes_form1} I^{(\nu)} \propto \sum_{i,\bk}|\bracket{\Phi^{\rm
N}_{0}}{\Phi^{\rm N}_{i,\bk}}{\bm{A}\cdot\sum_j{\hat{\bm{p}}_j}}|^2  
\delta (h\nu + E^{\rm N}_{0} - E^{\rm N}_{i,\bk})\,,
\end{equation}
which contains the squared modulus of the light-matter interaction operator in the dipole approximation --- 
where ${\bm A}$ is the amplitude of the semi-classical vector-potential 
and $\hat{\bm{p}}_j=-i\hbar\nabla_j$ is the linear momentum operator for the $j^{\rm th}$ electron,
Equation~\eqref{pes_form1} can be written in terms of single-particle Dyson orbitals
$\phi^{\rm d}_i(\br) = \bracket{\Phi^{\rm N-1}_{i}}{\Phi^{\rm N}_0}{\hat{\Psi}(\br)}$
and binding energies $E^{\rm b}_i=E^{\rm N-1}_i-E^{\rm N}_0$, as~\cite{walter_photoelectron_2008}:
\begin{equation}\label{pes_form3} 
I^{(\nu)} \propto \sum_{i,\bk} |\bracket{\phi^{\rm
d}_{i}}{\xi_{\bk}}{ \bm{A}\cdot \hat{\bm{p}}_i}|^2  \delta \robra{h\nu 
-E^{\rm b}_i
-\frac{\hbar^2 \bk^2}{2m}}.
\end{equation}
More details on the calculation of $I^{(\nu)}$ can be found in Ref.~\citenum{nguyen_first-principles_2015}. The excitation energy is now expressed in terms of the kinetic energy $\hbar^2 \bk^2/2m$ of the ejected electron and its binding energy $E^{\rm b}_i$ defined as the negative of the Dyson orbital energy $\varepsilon^{\rm d}_i$.

Dyson orbitals, whose energies are the poles of the one-body Green's function, fulfil 
(at least for discrete states~\cite{Onida2002}) the quasiparticle equation
\begin{equation}
  \label{eq:QP}
  \left[ \hat{T} + \hat{v} + \hat{\Sigma}(\varepsilon^{\rm d}_i) \right] | \phi^{\rm d}_{i} \rangle = 
     \varepsilon^{\rm d}_i \, | \phi^{\rm d}_{i} \rangle,
\end{equation}
where $\hat{v}$ is the sum of the external and Hartree potentials and $\hat{\Sigma}$ is the electron-electron self-energy. The calculation of Dyson orbitals should in principle be carried out within the framework of many-body perturbation methods~\cite{Onida2002}.
However, there is a strong analogy between~\eqn{eq:QP} and the generalized eigenvalue equation [\eqn{eq:KIPZ_eigenvalue_equation}] of Koopmans-compliant functionals, where the local and orbital density dependent operator $\hat{v}_i$ acting on the variational orbitals $|\varphi_i\rangle$ can be seen as a simplification of a non-local and frequency-dependent self-energy, as argued in Ref.~[\citenum{ferr+14prb}]. The canonical orbitals produced by an orbital-density-dependent calculation have in this perspective a natural interpretation as Dyson orbitals, and their energies as particle-removal energies. This justifies the choice to use them to construct photoemission spectra, in a framework which is computationally much less expensive than solving Eq.(\ref{eq:QP}) within state-of-the-art Green's function methods.

\begin{figure}
  \includegraphics[scale=0.3]{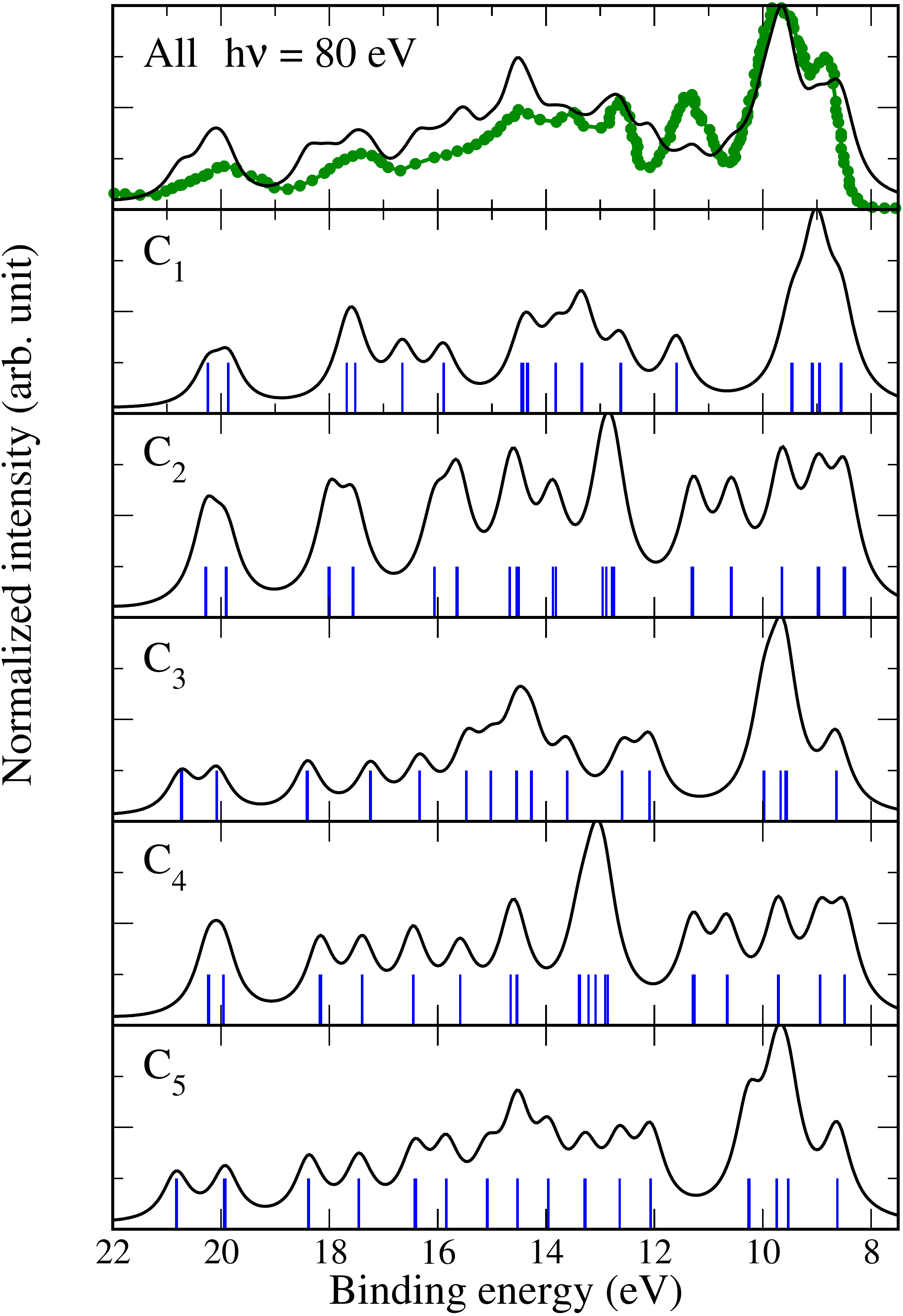}
  \caption{\label{ups_c_all} UPS spectra for cytosine tautomers calculated with the KIPZ functional (black line), and plotted as a function of electron binding energy ($h\nu-\hbar^2 k^2/(2m)$). The calculations are performed for an incoming photon energy of 80 eV. In the top panel we show the total theoretical spectrum (black line), which is averaged over the five most stable tautomers with Boltzmann weighing factors, and compared with experimental gas-phase UPS measurements (green dots) using radiation with the same photon energy~\cite{trofimov_photoelectron_2006}. The blue bars in each plot mark the energies of KIPZ canonical orbitals.} 
\end{figure}

\begin{figure}
  \includegraphics[scale=0.3]{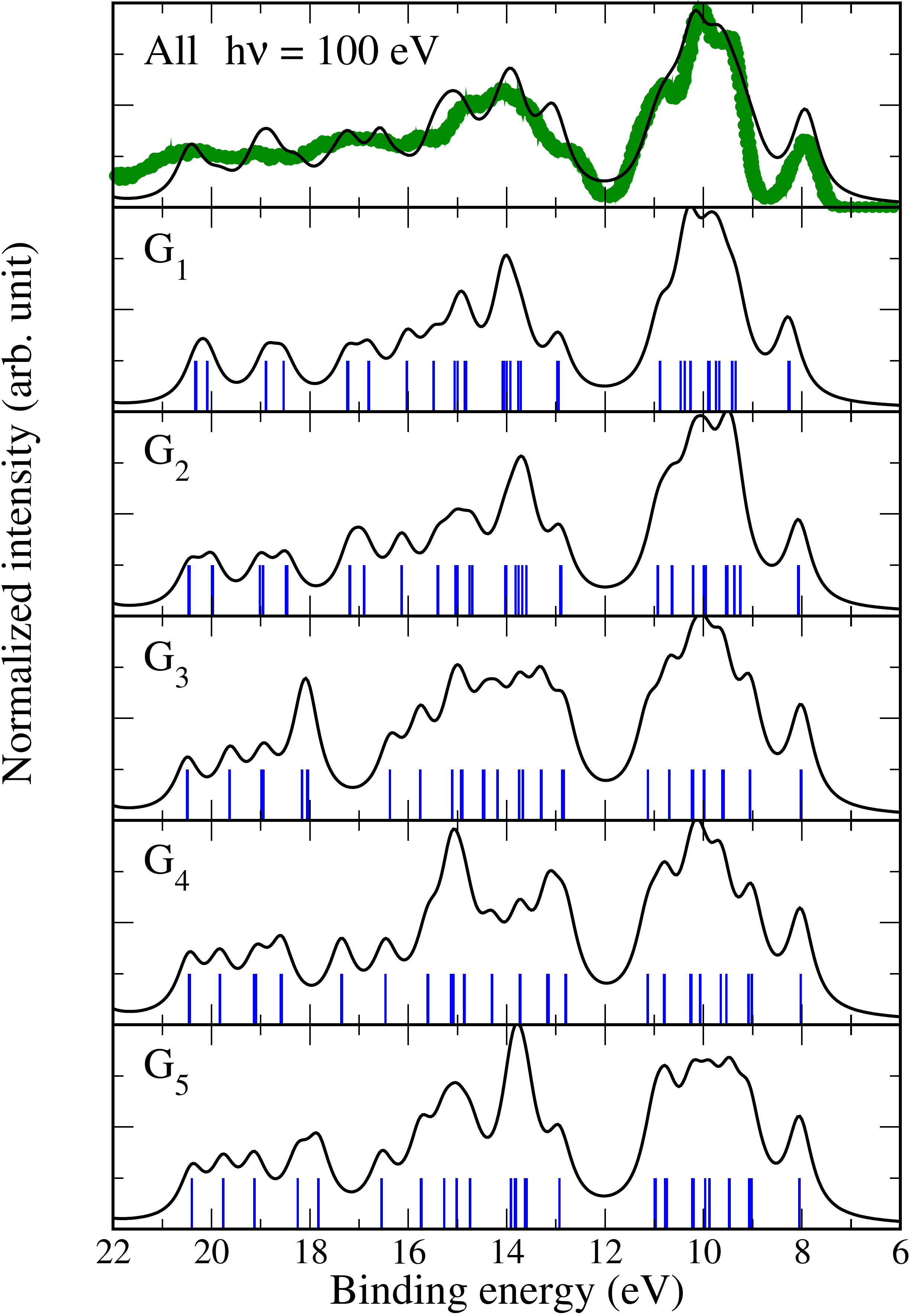}
  \caption{\label{ups_g_all} Same as \fig{ups_c_all} for guanine, and for an incoming photon energy of 100~eV. Experimental data are from Ref.~\citenum{zaytseva_theoretical_2009}.}
\end{figure}

The KIPZ results for nucleobases' spectra are shown in Figs.~\ref{ups_atu},~\ref{ups_c_all} and~\ref{ups_g_all}. Theoretical spectra are compared directly with experimental data only for the bases that have one single stable tautomer. This is not the case for G and C, for which the comparison is carried out between the experimental value 
and the Boltzmann average of the spectra of the five different tautomers. 
All photoemission spectra are computed with incoming photon energies $h\nu$ taken from the referenced experiments. For A, T, and C $h\nu = 80$ eV (Ref.~\citenum{trofimov_photoelectron_2006}), for G $h\nu = 100$~eV (Ref.~\citenum{zaytseva_theoretical_2009}), and for U $h\nu = 21.22$~eV (Ref.~\citenum{odonnell_ab_1980}). A detailed study of the effects of different photon energies on the shape and intensities of the photoemission spectra is shown in Figs. S1 and S2 in the Supporting Information~\cite{Supplemental_material}. 

The agreement between computed spectra and experiments is remarkable, not only for the peak positions but also for the shapes and intensities of the spectral peaks. 
The success of KIPZ can be explained through its ability to correct KS eigenvalues of approximate DFT by aligning them to particle removal energies through the Koopmans' condition, and by inheriting from PZ the property of being exact in the one-electron limit~\cite{Borghi_PRB_2014}. 
Such behavior is essential for the prediction of fundamental gaps and excitation energies~\cite{refaely-abramson_quasiparticle_2012}.
Another feature of KIPZ is its ability to modify not only the electronic excitation energies of approximate DFT, but also the manifold of electronic orbitals (i.e., the single-particle density-matrix)~\cite{Borghi_PRB_2014}. A change in the shape of single-particle orbitals, which results in a change of the one-body density-matrix of quasiparticles, 
affects both photoemission peak intensities and positions, ultimately affecting the accuracy of the simulated spectra.

The availability of experimental data from angle-resolved photoemission spectroscopy (ARPES) for DNA and RNA nucleobases enables us to comment also on the ability of KIPZ to predict the correct ordering, as well as the binding energy, of the orbitals close to the HOMO. 
The KIPZ results show that the type and order of the five orbitals with the lowest binding energies in the stable tautomers of A, T and U are the same, i.e., from HOMO to HOMO-4 the character of the orbitals is $\pi$, $\sigma$, $\pi$, $\sigma$, $\pi$ (see Fig. S3 in the Supporting Information~\cite{Supplemental_material} for a plot of the orbital densities in each of them),
in agreement with ARPES measurements and with other highly-accurate quantum chemistry calculations (using the full 
third-order algebraic diagrammatic construction 
[ACD(3)]~\cite{trofimov_photoelectron_2006} or the equation-of-motion coupled-cluster with single and double substitutions [EOM-IP-CCSD]~\cite{bravaya_electronic_2010}).
 
In the case of C and G, orbital ordering depends on the tautomer considered. The KIPZ orbital ordering which matches experimental ARPES measurements is the one of the most stable tautomer, i.e., the orbital ordering of C$_2$ ($\pi$, $\sigma$, $\pi$, $\sigma$ $\pi$) and G$_1$, respectively.
Other tautomers such as C$_1$, C$_3$ and C$_5$ have an orbital ordering in which the 2$^{\rm nd}$ and 3$^{\rm rd}$ orbitals are swapped with respect to C$_2$, while the ordering of G$_1$, G$_3$, G$_4$ and G$_5$ differs from G$_2$ by a swap of the 3$^{\rm rd}$ and 4$^{\rm th}$ orbitals. More details on the nature of the orbitals close to the HOMO can be found in Figs. S4 and S5 in the Supporting Information~\cite{Supplemental_material}.

It is worth to stressing that while for C and G the orbital character ($\sigma$ or $\pi$) and ordering appear to be determined from the most stable tautomers, the prediction of shape, intensity, and peak positions in the photoemission spectra requires the contribution of all tautomers, each weighted with its Boltzmann factor.
If we look in detail at the spectrum of each single tautomer of C or G (Figs.~\ref{ups_c_all} and~\ref{ups_g_all}), we see that from their overall appearance these can be divided into two groups. In the case of C, the spectra of C$_1$, C$_3$ and C$_5$ are similar, and are characterized by a low-energy prominent peak which appears in the Boltzmann-averaged spectrum on the first panel of Fig.~\ref{ups_c_all}. We can thus say that the electronic excitations of C$_1$, C$_3$ and C$_5$ are fundamental in determining the experimental spectrum of C, even though they do not include the most stable tautomer (C$_2$). Concerning C, we can determine a second group of tautomers with similar spectra, i.e., C$_2$ and C$_3$. The two groups of tautomers differ by the fact that a hydrogen atom is bound to an oxygen in the case of C$_2$ and C$_3$, and to a nitrogen in the case of C$_1$, C$_3$ and C$_5$ (see also \fig{fig:struc_properties}).
Correspondingly, one can observe similarities, which are less pronounced than in the case of C, in the spectra of G$_1$ and G$_2$. These two first tautomers can be grouped together and contrasted to G$_3$, G$_4$ and G$_5$, which have in common the presence of a hydrogen atom bound to an oxygen rather than a nitrogen.
All the above remarks emphasize how photoemission peaks and the overall photoemission spectra can in principle be used in order to extract information not only on the type of nucleobase, but also on the precise tautomeric forms present in a sample.

\subsection{Geometry optimization}\label{sec:geometry_results}

In this section we discuss the effects of structural optimization within the KIPZ framework on the electronic properties of DNA and RNA nucleobases. The self-consistent optimization procedure, involving atomic positions and screening coefficient $\alpha$, was outlined in Sec.~\ref{sec:screen_opt} and Fig.~\ref{fig:scf_flow}.
In Table \ref{tab_error_geos}, we test accuracy in computing intra-molecular bond lengths and angles by comparing the mean absolute error (MAE) with respect to experimental data of the structural predictions coming from several {\it ab initio} methods.
In~\rtab{tab_error_geos} we show data obtained by PBE, PZ, KIPZ, and PBE0.

Experimental data are taken from the Cambridge Structural Database by Clowney et al.~\cite{clowney_geometric_1996} where bond lengths are measured at room temperature through high-resolution X-ray and neutron diffraction; these data should be compared to the most stable tautomeric form of each nucleobase. In order to be consistent with these data, in~\rtab{tab_error_geos} we show the MAE with respect to experiment only for bond lengths and angles of those bases that have a single stable tautomer at room temperature, namely A, T and U. More detailed information about structural properties is available in Table S1, S2 and S3 in the Supporting Information~\cite{Supplemental_material}.

By looking at~\rtab{tab_error_geos}, one sees that the maximum error in predicting structural properties is displayed by DFT-PBE (MAE $> 1.39$\%). The relatively smaller accuracy of this functional can be partly explained from its self-interaction error, which results in a slightly increased spread of orbital densities. This can in turn affect geometrical properties, and in molecules it typically results in bond-lengths that exceed experimental values~\cite{Borghi_PRB_2014}. In contrast, the PZ functional, usually over-correcting the self-interaction error in molecules, tends to under-estimate bond lengths. The KIPZ functional, thanks to its {\it ab initio} screening factor, can interpolate between the two opposite behaviors of PBE and PZ, resulting in more accurate estimates of geometrical parameters. Our results show that the accuracy of KIPZ (MAE $\sim 0.6$\%) in computing bond lengths is better than PBE, and close to that of the PBE0 hybrid functional.
The same cannot be said for angles, for which the discrepancy between the different methods shown is much smaller, and shows no clear trend.
\begin{table}
\caption{\label{tab_error_geos} Relative mean absolute error of bond lengths and angles of A, T, and U molecules computed using PBE, PZ, KIPZ, and PBE0 methods, compared with experiments~\cite{clowney_geometric_1996}.}
\begin{ruledtabular}
\begin{tabular}{l r r r r r r }
    \% error & & PBE & PZ & KIPZ & PBE0  \\
\hline
              & A & 1.39 & 0.67 & 0.70 & 0.56 \\
$\bar{R}_{ij}$& T & 1.77 & 0.89 & 0.55 & 0.85 \\
& U &1.76 & 0.96 &0.70 & 0.93 \\
\hline
& A & 0.51& 0.52 & 0.56 & 0.54 \\
$\bar{\theta}_{ijk}$& T & 0.74 & 0.59& 0.64 & 0.63 \\
& U &1.05 &0.69 & 0.64& 0.81  \\
\end{tabular} 
\end{ruledtabular}
\end {table}
%

\begin{table}[!htb]
\caption{\label{tab_dihedral_angle} Absolute dihedral angles ($\beta_1$ and $\beta_2$) of the nucleobases with amino group. For A $\beta_1=\angle{\rm H_{2}N_{10}C_{4}N_{3}}$ and $\beta_2=\angle{\rm H_{1}N_{10}C_{4}C_{5}}$; for G$_{1..5}$ $\beta_1=\angle{\rm H_{2}N_{10}C_{2}N_{1}}$ and $\beta_2=\angle{\rm H_{1}N_{10}C_{2}N_{3}}$, for C$_{1,2,4}$ $\beta_1=\angle{\rm H_{1}N_{7}C_{4}C_{5}}$ and $\beta_2=\angle{\rm H_{2}N_{7}C_{4}N_{3}}$. The indices referring to atoms in nucleobase structures are explained in Fig. S6 in the Supporting Information~\cite{Supplemental_material}.}

\begin{ruledtabular}
 \begin{tabular}{l r r r r r r r}
      &     &$\beta_1$&     &  &     & $\beta_2$&     \\
       & PBE & KIPZ    & MP2\cite{doi:10.1021/j100063a019} &  & PBE & KIPZ     & MP2\cite{doi:10.1021/j100063a019} \\
 \hline 
 A     & 0.05    &23.16    &18.70 &  &0.05     &23.15     &21.10\\	
 G$_1$ &2.00     &10.80    &-    & &2.03      &46.12          &-   \\
 G$_2$ &2.05     &17.60    &11.80 &  &2.03     &35.36          &43.20  \\
 G$_3$ &1.82     &23.69    &-    &  &1.92     &24.44          &-  \\
 G$_4$ &1.81     &25.35    &-    &  &1.88     & 22.51         &-  \\
 G$_5$ &1.82     &21.97    &-    &  &1.88     & 27.69         &-  \\
 C$_1$ &8.25     &30.64    &26.20 &  &7.20     &17.27          &14.10  \\
 C$_2$ &8.68     &27.97    &-    &  &8.01     &21.01          &-  \\
 C$_4$ &8.90     &28.57    &-    &  &8.02     &20.53          &-  \\
\end{tabular}
\end{ruledtabular}
\end{table}

\begin{figure}
  \includegraphics[scale=0.32]{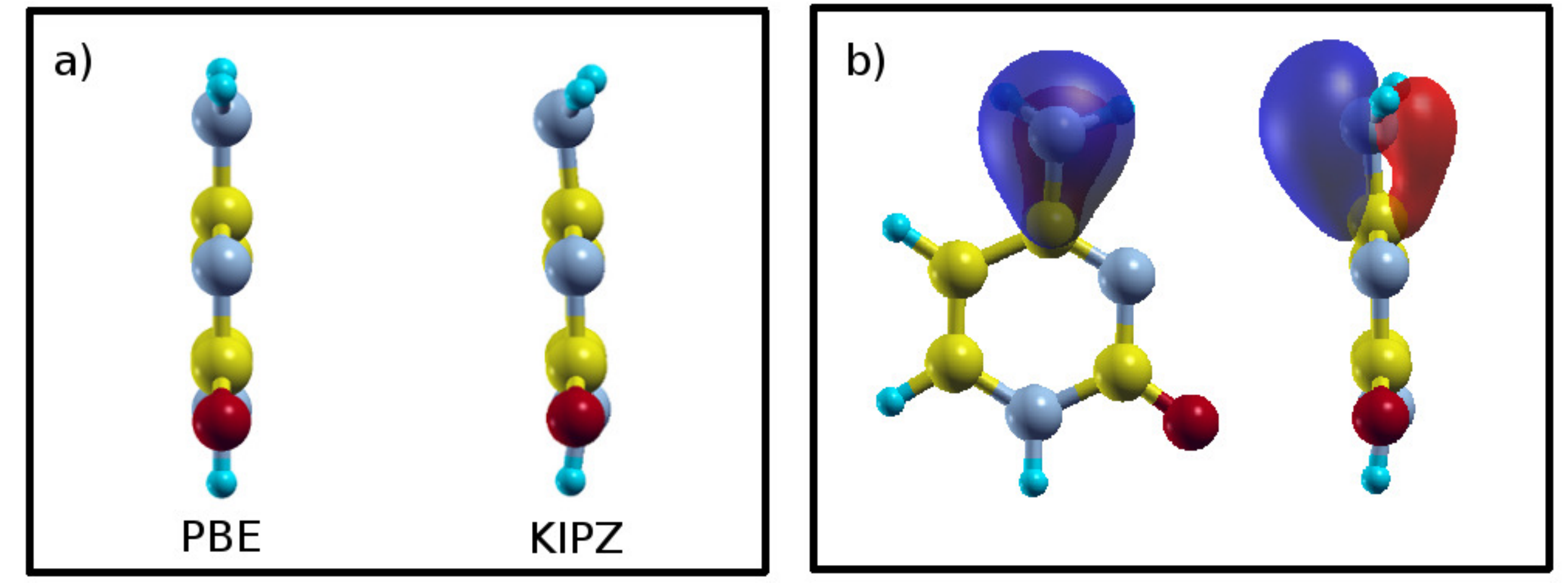}
  \caption{\label{iso_plot_evc0} (a) Side view of the C$_1$ structures optimized using PBE and KIPZ functionals. (b) Side view and top view of C$_1$, with an iso-surface plot of one of the variational orbitals supporting the bond of the amino group.} 
\end{figure}

 \begin{table}[!htb]
 \caption{\label{ip_ea_vs_planar} IPs and EAs of dipole-bound states computed using the KIPZ functional on top of PBE (@PBE) structure or on top of the structure resulting from scf-KIPZ optimization (@KIPZ, see Fig.~\ref{fig:scf_flow}).}
\begin{ruledtabular}
 \begin{tabular}{l c c c c c c c }
\hline
       &     &IP&     &  & &     EA&     \\
 geo.     & @PBE & @KIPZ & Expt.\footnote{\label{expt11}Collected from Reference~\onlinecite{roca-sanjuan_ab_2006}.} & & @PBE & @KIPZ & Expt.\footnote{\label{expt77}References~\onlinecite{desfrancois_electron_1996}
and~\onlinecite{schiedt_anion_1998}.} \\
 \hline 
 A     & 8.25 &   8.41    & 8.30$-$8.50 &  &-0.01 &-0.02& 0.01    \\	
 G$_1$ & 7.86 &   8.25    & 8.00$-$8.30 &  &0.08 & 0.03    &  $-$   \\
 G$_2$ & 7.77 &   8.07    & $-$         &  &0.13 &0.08     &  $-$   \\
 G$_3$ & 7.76 &   8.01    & $-$         &  &0.02 &-0.02 &  $-$   \\
 G$_4$ & 7.79 &   8.02    & $-$         &  &0.01 &0.01     &  $-$   \\
 G$_5$ & 7.78 &   8.05    & $-$         &  &0.04 &0.02 &  $-$   \\
 C$_1$ & 8.48 &   8.70    & 8.80$-$9.00 &  &0.11 &0.11 & 0.23   \\
 C$_2$ & 8.56 &   8.67    & $-$         &  &0.05 &0.03 & 0.09  \\
 C$_4$ & 8.57 &   8.67    & $-$         &  &0.03 &0.05 & $-$     \\
\end{tabular}
\end{ruledtabular}
\end{table}

A remarkable success of the structural optimizations done with the KIPZ functional is the correct description of the slight tilt of the amino groups of nucleobases with respect to their aromatic rings, which can be seen in A, G$_1$, G$_2$, G$_3$, G$_4$, G$_5$, C$_1$, C$_2$ and C$_4$. A correct prediction of this nonplanarity is an important step towards accurate predictions of the structure of DNA and towards the understanding of molecular recognition processes in biological systems.
The presence of the amino-group tilt was subject to some controversy in the past 
(for a detailed discussion, see, e.g., 
Refs.~\citenum{hobza_structure_1999} and \citenum{sponer_dna_1996}).
Theoretical studies of DNA and RNA nucleobases~\cite{riggs_ab_1991, brown_study_1989} using the HF method with the 3-21G basis set originally suggested the molecules to be perfectly planar. Subsequent calculations carried out at the HF level with polarized basis sets of atomic orbitals were instead able to observe a weak nonplanarity of the amino groups of the base molecules~\cite{leszczynski_are_1992}. Later, post-HF approaches indicated an even stronger amino-group pyramidalization.
For C$_1$, for example, Bludsky et al~\cite{bludsky_amino_1996} obtained amino group hydrogen dihedral angles of 5.5$^{\rm o}$ and 21.4$^{\rm o}$ using the HF/6-31G$^{**}$
and MP2/6-31G$^{*}$ levels of theory, respectively. This MP2 result is close to the predictions by Sponer and Habza of a dihedral angle of 26.2$^{\rm o}$~\cite{doi:10.1021/j100063a019}. 
Our DFT-PBE calculations, in agreement with those by Di Felice et al~\cite{di_felice_ab_2001}, do not find significant deviations from planarity (see~\rtab{tab_dihedral_angle}), with a dihedral angle of 8.25$^{\rm o}$ for C$_1$, and even smaller angles for the other DNA bases.
The strong amino-group pyramidalization can be instead reproduced very well within KIPZ, with amino hydrogen dihedral angles of the same order of the MP2 results. 
On panel (a) of~\fig{iso_plot_evc0}, we show as an example the equilibrium geometry of the C$_1$ molecule predicted by KIPZ, as compared to the PBE structure. On panel (b) we show the KIPZ tilted structure with the variational orbital building the distorted $\pi$-bond that supports the connection between the amino group and the molecule. The orbital asymmetry shown is correlated to a switch from a {\it sp$^2$}-like hybridization of the N atom to a more {\it sp$^3$}-like bond configuration in the tilted structure.

We find that predicting correctly the amino group shape of the nucleobases is important in view of computing accurately the electronic excitation spectrum of the molecules. 
In~\rtab{ip_ea_vs_planar} we compare IPs and EAs computed with the KIPZ functional top of the PBE structure (a mostly planar molecule) and on top of the KIPZ (nonplanar) structure. Describing well the pyramidalization of NH$_2$- group increases ionization energies by about 0.2$-$0.3 eV (with a much less pronounced effect on LUMO states), resulting in theoretical predictions that are closer to experiments. We stress that all the results of Sec.~\ref{sec:technical_details},~\ref{sec:binding_energies}, and~\ref{sec:ups} used this scf-KIPZ approach for determining geometry.
\section{Conclusion}

In this paper we have explored the capability of the KIPZ functional to predict both spectral properties, such as ionization potential, electron affinities, ultraviolet photoemission spectral, and geometries of DNA and RNA nucleobases and their tautomer variants, 
showing an excellent agreement with experiments, with mean absolute errors for the first IPs and EAs that are smaller than 0.1 eV. The accuracy of the KIPZ functional in predicting IPs and EAs of nucleobases is comparable to that of more computationally intensive methods derived from many-body perturbation theory, such as G$_0$W$_0$ and scf-GW,
or quantum chemistry, such as CASPT2 and CCSD(T). 
In addition, for EAs the empty excited states of the nucleobases can be found in two variants: the delocalized and weakly bound DB states and the localized and unbound VB states, which are close in energy, and which make the study of these systems with localized basis set extremely challenging, but yield very accurate results in the case of KIPZ.

Similarly, the photoemission spectra of nucleobase molecules show an excellent 
agreement with UPS data measured at the same incoming photon energy. These results support the suggestion (see Refs.~\citenum{ferr+14prb} and~\citenum{nguyen_first-principles_2015}) that KC functionals can be seen as a beyond-DFT approach where 
the spectral potential~\cite{Gatti2007prl}, rather than the exchange-correlation one, is directly approximated, and provide both a conceptual and a practical framework to predict spectral properties from functional theories, rather than perturbative approaches. In addition, the excellent agreement between theoretical and experimental spectra allows us to assign the orbitals of the low binding energy UPS excitations using the KIPZ eigenstates, as well as to resolve the experimental spectra of C and G molecules by attributing them to a weighted spectrum of their tautomers. 

By exploring self-consistent screening (scf-KIPZ) to optimize molecular geometries we find that correctly predicting the structural properties of the bases, especially for the amino groups, yields a better agreement between theoretical and experimental IP energies. Overall, we believe that our results are a step towards further studies of the electronic structure 
of complex DNA and RNA sequences, for which methods from many-body perturbation theory or quantum chemistry would be computationally very challenging. 
 \begin{acknowledgments}
We acknowledge partial support from the Swiss National Centre
for Computational Design and Discovery of Novel Materials (MARVEL), and the EU Centre of Excellence "MaX - Material design at the eXascale" (Grant No. 676598).
\end{acknowledgments}

\bibliography{biblio}
\bibliographystyle{aip}
\end{document}